\documentclass[journal,article,submit,pdftex,moreauthors]{Definitions/mdpi} 

\renewcommand{\linenumbers}{}

\newcommand{\actaa}{\mbox{\it Acta Astronomica}}
\newcommand{\aj}{\mbox{\it Astronomical J.}}
\newcommand{\apj}{\mbox{\it Astrophys. J.}}
\newcommand{\apjl}{\mbox{\it Astrophys. J.}}
\newcommand{\apjs}{\mbox{\it Astrophys. J.}}

\newcommand{\aap}{\mbox{\it Astron. Astrophys.}}

\newcommand{\araa}{\mbox{\it Annu. Rev. Astron. Astrophys.}}

\newcommand{\mnras}{\mbox{\it Mon. Not. R. Astron. Soc.}}
\newcommand{\nar}{\mbox{\it New Astronomy Rev.}}
\newcommand{\nat}{\mbox{\it Nature}}

\newcommand{\physrep}{\mbox{\it Phys. Rep.}}
\newcommand{\pasj}{\mbox{\it Publ. Astr. Soc. Japan}}

\newcommand{\prd}   {{\it Physical Review D}}
\newcommand{\prl}   {{\it Physical Review Letters}}

\firstpage{1} 
\pubvolume{1}
\issuenum{1}
\articlenumber{0}
\pubyear{2024}
\copyrightyear{2024}
\datereceived{ } 
\daterevised{ } 
\dateaccepted{ } 
\datepublished{ } 
\hreflink{https://doi.org/} 

\Title{Gamma-ray bursts: what do we know today that we did not know 10 years ago?}

\TitleCitation{Title}

\Author{Asaf Pe'er $^{1}$\orcidA{}}

\AuthorNames{Asaf Pe'er}

\AuthorCitation{Pe'er, A.}

\address{%
$^{1}$ \quad Department of Physics, Bar Ilan University,\\
Ramat-Gan 52900, Israel;  asaf.peer@biu.ac.il}

\abstract{I discuss here the progress made in the last decade on few of the key open problems in GRB physics. These include: (1) the nature of GRB progenitors, and the outliers found to the collapsar/merger scenarios;  (2) Jet structures,  whose existence became evident following GRB/GW170817; (3) 
the great progress made in understanding the GRB jet launching mechanisms, enabled by general-relativistic magneto-hydrodynamic (GR-MHD) codes; (4) recent studies of magnetic reconnection as a valid energy dissipation mechanism; (5) the early afterglow, which may be highly affected by a wind bubble, as well as recent indication that in many GRBs, the Lorentz factor is only a few tens, rather than few hundreds. I highlight some recent observational progress, including major breakthrough in detecting TeV photons and the on-going debate about their origin, polarization measurements, as well as the pair annihilation line recently detected in GRB 221009A, and its implications on the prompt emission physics. I point into some open questions that I anticipate would be at the forefront of GRB research in the next decade.
}


\begin{document}


\section{Introduction}
\label{sec:into}

Since the realization in the early 1990's that gamma-ray bursts (GRBs) are of cosmological origin, and therefore release a huge amount of energy - typically $10^{50}-10^{53}$~erg in the form of
gamma-rays over a few second duration, they have challenged the boundary of physics and have fascinated the imagination of many astronomers. They further show an extremely rich
phenomenology, with broad band spectra that extends over the entire electromagnetic spectrum - from radio to the TeV range, and with highly variable lightcurve (down to 0.01 s in some cases).
As no two GRBs are identical, the data challange both observational campaigns and theoretical models (see, e.g., \cite{Peer15, KZ15, Zhang18, Bosnjak+22} for recent reviews). 

Although naturally the field is maturing, interestingly, many 
fundamental questions about the underlying nature and physics of GRBs still remain unsolved. In fact, there are many open problems whose answers were unknown, or, for the least, not in a consensus 10 years ago, and still are today. Nonetheless, clearly, there had been a huge volume of works (according to NASA ADS, about 18,000 papers whose title contain "GRB" appeared since 2014), which addressed some of the questions - while opening new ones.   
It is therefore useful to look at the big open questions that had occupied researchers in this field, and the progress that was made in the past decade. This enables one to look into the future, and foresee the progress that we hope and expect to make in the coming years.

Phenomenologically, GRBs are traditionally classified into two
categories: the ``short'' (or short/hard) GRBs, with an average
duration of a few tenths of a second; and the ``long'' (or long/soft)
GRBs, with an average duration of 20~s. The dividing line is typically
found at $\sim 2$~s (\cite{Kouveliotou+93, Paciesas+99, Gruber+14,
  VonKienlin+14, Lien+16}). Furthermore, there are evidence for a sub-class of "ultra-long" GRBs \cite{Gendre+13, Levan+14}, whose duration exceeds 10,000 seconds. It is unclear whether these "ultra-long" GRBs form a separate population with distinct physical properties or not. For example, the ultra-long GRB220627A did not show any different properties than long GRBs (in terms of jet break or ambient density) \cite{DeWet+23} supporting the idea that its progenitor is similar in nature to other, more standard long GRBs. 
  
Despite this huge variation in the GRB duration, as well as the X- and gamma-ray lightcurves: with some GRBs being very "spiky" and others show a much smoother lightcurve, all GRBs share some common spectral features. These include a distinct observed spectral peak at sub-MeV energy range, and spectra that is often modeled (even if cruedly, see below) as a "broken power law" \cite{Band+93} with distinct low and high energy spectral slopes. In recent years, there are evidence for an additional high energy component, that extends, in some cases, to the TeV range \cite[e.g.,][]{Acciari+19, HESS19}. Furthermore, clearly, all GRBs are transients, lasting a (relatively) short duration, and no repetition has ever been found.

\begin{figure}
\includegraphics[width=11cm]{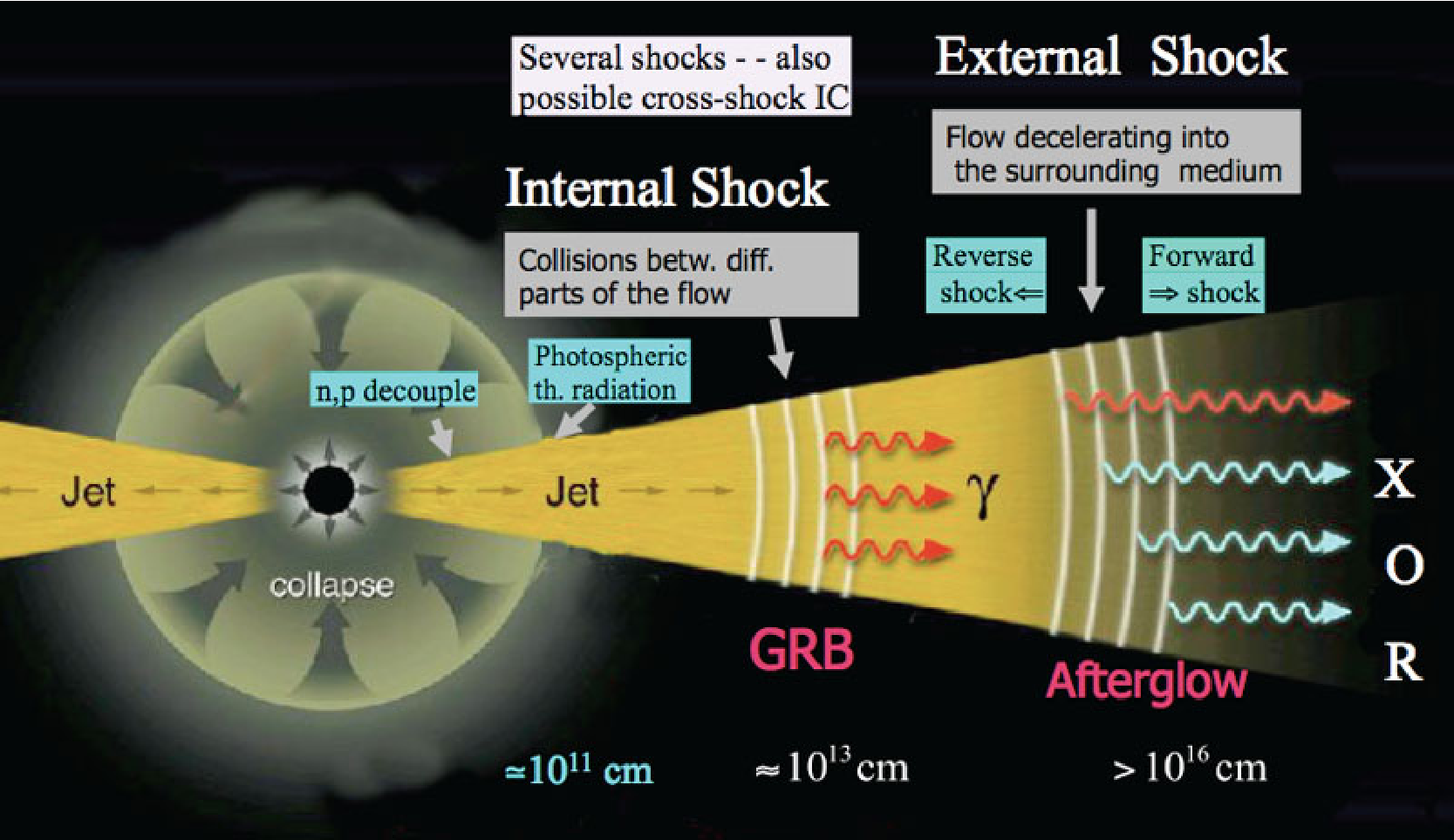}
\caption{Illustration showing the basic ingredients of the GRB
  ``fireball'' model. (1) The source of energy is a collapse of a
  massive star (or binary merger, not shown here). (2) Part
  of this energy is used in producing the relativistic jet. This could
  be mediated by hot photons (``fireball''), or by magnetic field. (3)
  The thermal photons decouple at the photosphere. (4) Part of the jet
  kinetic energy is dissipated (by internal collisions, in this
  picture) to produce the observed $\gamma$ rays. (5) The remaining
  kinetic energy is deposited into the surrounding medium, heating it
  and producing the observed afterglow. Figure is taken from
  \cite{MR14}.  }
\label{fig:fireball}
\end{figure}

Interestingly enough, the basic skeleton of the theoretical understanding of GRBs had not changed: it is the famous GRB "fireball" model   \cite{Pac86, Goodman86, RM92, MLR93, RM94}; see Figure \ref{fig:fireball}. 
According to the ``fireball'' model, a vast explosion, associated with
the formation of a black hole, leads to the ejection of material that
forms a relativistic jet. There is still a debate about the role played by the magnetic field in the jet acceleration process (see below). At a second stage, some of the jet kinetic
energy is dissipated, e.g., by internal shock waves (\cite{RM94, PX94, SP97, Kobayashi+97, DM98} and many others),
magnetic reconnection (e.g., \cite{Thompson94, Katz97, MR97b, MU12, Cerutti+13b} and many more) or other,
unspecified mechanism, producing the observed prompt emission signal. Following the dissipation,
the material in the jet continues to expand relativistically
into the interstellar medium (ISM), driving a relativistic shock
wave. This shock wave, in turn, heats the ISM material, generates a magnetic field, and accelerates a substantial fraction of the particles to a non-thermal distribution above the thermal peak. During the acceleration process, these non-thermal particles acquire a power law energy distribution. They then radiate synchrotron emission, which is the main source of the afterglow emission \cite{WRM97, SPN98}. This afterglow is routinely detected since 1997
(\cite{VanPar+97, WRM97, Galama+98}).

One of the open questions that is still debatable is the role of magnetic fields in this process. The original "fireball" model assumed a baryonic-dominated outflow \cite{CR78, Pac86, Goodman86, SP90, Pac90}, in which the gravitational energy is converted to the jet kinetic energy via neutrino-anti-neutrino annihilation, that produces a copious number of $e^\pm$ pairs. The "fireball" is thus composed of photons, pairs, as well as electrons and baryons (that carry the momentum), relics from the explosion that initiated the process. However, already in the 1990's it was suggested that magnetic fields may play a significant role in shaping the dynamics of the flow
\cite{Usov92, Thompson94, MR97b, SDD01, Drenkhahn02, DS02, SD04, LB03, Komissarov+09}. According to the "magnetized fireball" models, most of the energy is carried in the form of Poynting flux, which is later used in accelerating and heating the gas via magnetic reconnection process. Although many details are uncertain (see below), this model has several advantages over the classical "fireball" model, and is thus considered by many as a viable alternative. However, this is still highly debatable.

The question of the role played by the magnetic field is related to the nature of the explosion that triggers this chain of events. Indeed, the nature of the explosion itself is also 
somewhat uncertain. The two leading models are the collapse of a
massive star (the so-called ``collapsar'' model; \cite{Woosley93,
  MW99, MacFadyen+01}), and merger of binary stars (\cite{Eichler+89,
  Narayan+92}), e.g., a neutron star (NS) and a black hole
(BH). Evidence for the connection of massive star collapse with GRBs
exist for 20 years, since the discovery of absorption lines in the
afterglow of GRB030329 (\cite{Hjorth+03, Stanek+03}). However, as I will discuss below, in recent years several outliers were found to this seemingly clear picture.

As for the origin of short GRBs, for many years only indirect evidence
connected them with the merger of binaries. Such evidence included
their observed occurrence location in the outskirts of their host
galaxies (\cite{Bloom+02, Gehrels+05, FB13, Berger14}), their existence in both star forming and elliptical galaxies, lack of association with supernovae, and their tendency to trace under-luminous locations within their host galaxies \cite{Fong+10}; see, e.g., \cite{Berger11}  for a review. This changed in
2017, with the discovery of gravitational waves associated with GRB170817
(\cite{Kasliwal+17, Abbott+17a, Abbott+17c}). Such gravitational wave signal, which
preceded this GRB by 1.7~s, could only originate from the merger of
two neutron stars.  This discovery, thus, served as a smoking gun
proving the merger origin of this GRB. This light-shedding event triggered an enormous amount of observations as well as theoretical modeling, significantly promoting our knowledge of both GRB progenitors and their jet properties, as is discussed below. 
I do point out, however, that this GRB was peculiar
in several ways, including being $10^2-10^4$ times less luminous
than typical short GRB. This may challenge the claim of
universality.

A unique observational consequence of this binary merger scenario is the rapid
production of heavy nuclei. The very high temperatures during the merger
event, $\gtrsim 10^{13}~{}^\circ$K, implies a release of a huge number of
neutrons and protons. These recombine into $\alpha$-particles, which
merge with free neutrons to assemble heavy seed nuclei (with nuclei
larger than the Iron nucleus). These heavy nuclei then radioactively
decay very rapidly, producing bright emission on a time scale of
$\sim$~day, which became known as ``kilonova''. The origin of the name is that it is approximately 1000 times brighter
than regular novae (\cite{Metzger+10, MB12}). The discovery of a kilonova was a major success of this model \cite{Kasen+17}. Nonetheless, here too several outliers were recently found, whose origin is still a mystery. I will briefly discuss these below.

Although the basic picture of GRB formation and evolution - namely, the 'skeleton' of the fireball model is well accepted by the
community, many, possibly most of the details remain open questions,
even after three decades of extensive research. GRBs represent a very
complicated environment, in which several independent processes of
energy exchange exist: from gravitational energy release that leads to
an explosion, through conversion of the explosion energy to kinetic
energy in the form of relativistic jet, and finally kinetic energy
dissipation leading to the observed radiative signal. Many of the
details of each of these processes is still actively debated in the
literature. The main reason for that is that we only see the final
outcome of this complicated chain of processes - the radiative signal
(spectra and lightcurve), which vary from burst to burst, from which
we try to deduce the full physics of the entire chain,

In the past decade, a plethora of data continuously streams, due to a large number of telescopes that are active at all wavelengths. One interesting consequence, is that in recent years several unique bursts have been detected, which challenge what is already considered as ``common knowledge'' in this field. Furthermore, the continuous flow of data serves as a great platform, on which new theoretical ideas flourish. 

Here, I highlight some of the big open questions, and discuss the progress made in the past decade, with a view into the next decade. Clearly, any such list is subjective; nonetheless, I believe that the questions I highlight here are representatives of at least the main stream in this field today. The big problems that I would like to highlight can be summarized as followed. 
\begin{enumerate}
\item The nature of the progenitor, or: what causes a GRB in the first place? As discussed above, both the "collapsar" and binary merger are considered valid scenarios. However, many of the details of the processes are still unknown, as well as the basic question of why two such different scenarios lead to similar observational outcome. Moreover, as we have seen in recent years, several observations challenge the simplified, binary picture. 
  
\item The jet composition and the origin of the magnetic field. While in the classical GRB "fireball" model \cite{RM92}, the jet is accelerated by the radiative pressure, today there are indications that the jet may be highly magnetized  \cite{Metzger+11, Bucciantini+12}, in which case the Blandford-Znajek process \cite{BZ77} may play a key role in the jet formation as well as its properties. 

\item The geometrical jet structure, namely its velocity profile, its dynamics and evolution. While early works considered, by large, a 'top-hat' jet, namely neglected a lateral jet profile, in recent years it became evident that GRB jets have a lateral structure, namely $\Gamma = \Gamma(\theta)$ \cite{ZWM03, PL23}. This affects the observed signal, which will be different for observers located at different angles to the jet axis. Furthermore, as of today there is still no clear understanding why GRB jets reach extremely high Lorentz factors, with $\Gamma$ of several hundreds in some cases, while other astronomical transients are "only" mildly relativistic at most. 

\item The nature of the energy dissipation mechanism that leads to the observed prompt emission signal (flux, spectrum and lightcurve) is unclear. Many early works considered internal shocks as the key mechanism for kinetic energy dissipation \cite{RM94, SP97, Kobayashi+97}. However, it was quickly realized that this process suffers from low energy conversion efficiency. Therefore, alternative models were suggested, such as magnetic reconnection \cite{MU12, WU17} or dominated contribution from the photosphere \cite{PMR06}.  

\item The radiative processes that lead to the resulting radiative signal, as well as other counterparts, such as cosmic rays, neutrinos or gravitational waves are still debatable. Production of the observed signal is the final outcome of a chain of events: from the dynamics, through energy conversion that accelerates particles to the radiative process. Many details of these events are uncertain, and therefore there is a high theoretical uncertainty in the expected signals.

\item The ambient medium density profile. The vast majority of early models considered the relativistic jet to expand into an either "constant density" environmental profile ($n(r) \propto r^0$) or, alternatively, a "wind" profile of the form $n(r) \propto r^{-2}$, resulting from a stellar wind at constant velocity (e.g., \cite{CL99, PK02}). However, in recent years there are increasing evidence that these approximations are too simplified. One naturally expects a 'wind bubble' structure around GRB progenitor stars \cite{Weaver+77, Dai04, VanMarle+08}. Such 'wind bubbles' result from stellar mass ejection prior to its terminal explosion that leads to a GRB, and can clearly affect the observed signal during the early afterglow phase \cite{PW06, Bucciantini+07, YD07}.

\item Finally, I will mention as open questions the connection of GRBs with other objects of interest, such as stellar evolution, star formation, host galaxies, supernovae, binary stars, etc. which are not yet fully understood. Similarly, the connection between GRBs and fundamental physics, such as the use of GRBs as cosmological probes \cite{Amati+02, Yonetoku+04}, Lorentz violation \cite{Abdo+09c}, etc. is a field that is still being explored.
\end{enumerate}

Clearly, all these questions are inter-related to each other, and answering one can provide important clues on the nature of others. However, interestingly enough, as of today there is no firm answer to any of these questions. 
Here, I highlight some of the recent developments that occurred in the past decade on addressing some of these questions, and try to predict where the next steps will be heading.
 
\section{The nature of GRB progenitor}
   
Already in the mid- 1990's, it became evident that GRBs are composed of two separate populations: the ``short'' (or short/hard) GRBs, with an average
duration of a few tenths of a second; and the ``long'' (or long/soft)
GRBs, with an average duration of 20~s. The dividing line is typically
found at $\sim 2$~s \cite{Kouveliotou+93, Paciesas+99}. While initially, many theoretical ideas were proposed to explain these results, theoretical models quickly converged into two: collapse of a massive star \cite{Woosley93,
  MW99, MacFadyen+01}, and merger of binary stars \cite{Eichler+89,
  Narayan+92}. 
 
\subsection{Long GRB progenitors: outliers to the "collapsar" model}  
In the early 2000's, direct evidence for the connection of massive star collapse with GRBs emerged, with the discovery of absorption lines in the
afterglow of GRB030329 \cite{Hjorth+03, Stanek+03}. Such absorption
lines are typical for core collapse supernova (known as SN type Ib/c \footnote{Supernova type Ib/c are core collapse supernovae with stripped hydrogen envelopes.}),
and their existence is a clear indication that the generation of
long GRBs is associated with a core collapse of a massive star. Following these discoveries, this became a 'common knowledge' \cite{WB06, Campana+06}. 

However, in recent years, evidence are accumulating that the picture is more complicated. Some GRBs that are clearly categorized as "long" GRBs, such as GRB211211A at redshift of $z=0.08$ ($T_{90} \sim 40$~s) \cite{Rastinejad+22, Troja+22, Yang+22} or GRB230307A at $z=0.065$ with $T_{90} \approx 45$~s \cite{Sun+23, Dichiara+23} (which is the second brightest GRB ever detected) are bright enough and close enough to show evidence for a supernova. However, instead of detecting a supernova as expected, both these GRBs show evidence for a kilonova, as expected from a binary merger.

For example, a clear evidence for kilonova emission was observed in GRB230307A, 
starting 2.4 days after the burst and lasting more than two months later
\cite{Yang+24}. Such a signal is a clear indication for a binary
neutron star (BNS) merger origin, and is not expected when collapse of
a massive star generates the GRB. Thus, this GRB challenges the
accepted division between the long and short GRB populations, as well
as the common belief that long GRBs originate from collapse of a single,
massive, star. Moreover, evidence for heavy element, such as Tellurium, generated by r-process were reported in the JWST lightcurve of this burst at late times, after 29 and 61 days \cite{Levan+24}. These results therefore provide a strong indication in favor of BNS merger progenitor for this GRB.  

Similar considerations led to the suggestion (\cite{Zhong+23})
that GRB211211A originates from a binary merger. 
Evidence for neutron star merger origin in GRB211211A was excess of
optical and near-infrared emission, both consistent with the kilonova
observed after the gravitational wave detected GW170817. However, it was argued
(\cite{BM23}) that an unusual collapsar could explain both the
duration of GRB 211211A and the r-process powered excess in its
afterglow. These GRBs therefore are either outliers to the long GRB populations, or possibly hint towards a new type of GRB progenitor.

\subsection{Binary merger as short GRB progenitors: final word?}
As opposed to the question of long GRB progenitors, prior to 2017 evidence for the association of short GRBs with binary mergers were only indirect. These include: (1) the association of short GRBs with elliptical galaxies  \cite{Berger14} as opposed to long GRBs which are associated with star forming galaxies; (2) the location of short GRBs within their host galaxies are observed at an offset from the galactic center \cite{Bloom+02, FB13}; (3) the lack of evidence for a supernova association; and (4) their location relative to the light: long GRBs are far more concentrated in the very brightest regions of their host galaxies  \cite{Fruchter+06} than short ones \cite{Fong+13}. 

This situation had dramatically changed in August 2017, with the association of the short GRB170817A to a gravitational wave source GW170817 \cite{Abbott+17a, Abbott+17b, Abbott+17c, Goldstein+17,  Kasliwal+17}. Since gravitational waves at the observed magnitude can only originate from binary neutron star mergers, this discovery served as a smoking gun for the association of short GRBs with the merger scenario. 
Furthermore, several hours later, an optical counterpart was discovered  with a luminosity, thermal spectrum, and rapid temporal decay consistent with those predicted for "kilonova" (KN) emission \cite{Tanvir+17, Cowp+17, Drout+17, Arcavi+17, Nicholl+17}. This emission is powered by the radioactive decay of heavy elements synthesized in the merger ejecta \cite{LP98, Metzger+10, MB12}. This discovery did not only confirm the NS merger origin of this burst, but also used to prove that the origin of heavy elements is indeed in the mergers of binaries, as long thought. 

Interestingly enough, signs of kilonova were observed earlier, 
associated with the short GRB 130603B \cite{Tanvir+13, Berger+13}, 
although no gravitational wave signal was detected from this event, as LIGO was not sensitive enough to detect a potential signal at that time. These results clearly indicate the merger origin of at least some of the short GRBs.

The picture, though, at least to me, is not complete yet. First, GRB170817, though clearly a light-shedding event, was a very peculiar GRB. 
In particular, its luminosity was two to four orders of magnitude lower than typical for short GRBs \cite{Fong+17}. Furthermore, as of today, this is still only a single, unique event. No other events were detected, although the prospects for additional detection in LIGO O3 run were good \cite{Howell+19}. It was argued, though, that this lack of additional detections can be used to constrain GRB jet opening angles, which are typically a few degrees \cite{Kapadia+24}.  

Thus, at least to my view, there is still a way to go before claiming that all short 
GRBs originate from a binary merger. In fact, there is at least one reported case, namely GRB 200826A with duration (T90) of $1.14 \pm
0.13$ seconds in the 50–300 keV energy range, which show clear indication for a collapsar progenitor   \cite{Ahumada+21}. This may very well be at the edge of the Gaussian distribution of long GRB duration, though its existence indicates that the categorization of GRBs need to be looked at on a case by case basis.   

\subsubsection{Lessons from GW/GRB170817A}

Despite the fact that, at least to my view, the final word about the origin of short GRBs had not been said yet, there is no doubt that GW/GRB170817A was a light-shedding event. It is therefore useful to briefly summarize the key lessons learned from this event. 
\begin{enumerate}
\item There is a clear association of (at least some) short GRB with binary neutron star merger.
\item The detection of kilonova: theoretical modeling shows that the matter that is expelled in the violent merger of two neutron stars can assemble into heavy elements such as gold and platinum in a process known as rapid neutron capture (r-process) nucleosynthesis. The radioactive decay of isotopes of the heavy elements is predicted to power a distinctive thermal glow (a ‘kilonova’;  \cite{Metzger+10, MB12}).The data confirms these predictions \cite{Kasen+17}.
\item Furthermore, the data constraints the maximum neutron star mass to be   $2.17 M_\odot$ \cite{MM17, Rezzolla+18}.
\item There is a clear evidence that the merger produces a relativistic jet \cite{Rezzolla+11} which is detected at late times \cite{PNR13}.
\item Late time observations revealed that the jet associated with GRB170817A is (i) structured; and (ii) viewed off axis (see section \ref{sec:structure} below).   
\end{enumerate}
 
Thus, there is no doubt that this was the single most important event in the history of GRBs, in terms of the information and insight it provided the community with.
 
\subsection{Magnetar giant flare: a distinct GRB population ?}

While the vast majority of GRBs are associated with a one-time terminal event, it had been suggested that some fraction may be associated with a repeated event.

Soft gamma repeaters (SGRs) are galactic X-ray stars that emit numerous short-duration (about 0.1~s) bursts of hard X-rays during sporadic active periods. They are thought to originate from magnetars, which are strongly magnetized neutron stars with emission powered by the dissipation of magnetic energy. Several SGRs have been detected in our galaxy \cite{Kouveliotou+94, VanderHorst+10, Kaneko+10}.
Most importantly, several giant flares have been detected from these magnetars \cite{Hurley+99, Mazets+99, Palmer+05}, with isotropic energy exceeding $10^{46}$~erg \cite{Palmer+05}. If such a magnetar is extra-galactic, the giant flare would be visible as a single flare, as lower energy flares are below current detectors limit. It was therefore proposed that some observed single-pulse short GRBs may be associated with these extragalactic magnetar giant flares (MGF) \cite{Pac92, Hurley+05, Svinkin+21}.   

Analysis showed that these MGF can account for a small fraction, of a few \% of the short GRB population \cite{Ofek07, Svinkin+15}. Nonetheless, as they represent an alternative channel for producing GRBs, and the only one that can lead to a repeater, they gain interest in recent years. Indeed, a recent analysis identified several nearby short GRBs that are unambiguously associated with MGF  \cite{Burns+21}.
Furthermore, this fraction depends on the detector's sensitivity: as MGFs are weaker than binary merger signal, yet they are more abundant than merger rate, this fraction is expected to grow in the future, when more sensitive instruments become available \cite{Beniamini+24}.

\section{Jet structure}
\label{sec:structure}

The fact that GRB explosions lead to the formation of jets (rather than spherical explosions) is well known for over 20 years, following observations of jet breaks \cite{Frail+01, Berger+03a, Racusin+09}. Indeed, such jet breaks are useful for GRB calorimetry \cite{Frail+01, PK02, Racusin+09, WWF11}, from which constraints on progenitor models can be derived, as well as the true GRB occurrence rates. For example, measuring 
the jet opening angle of 29 short GRBs \cite{Rouco+23} enabled to calculate the true event rates. The inferred rates ($\sim 1000~{\rm Gpc}^{-3}~{\rm  yr}^{-1}$) are consistent with the rate of NS-NS mergers, but are higher than BH-NS merger rate \cite{OK10, Grunthal+21} by a factor of 2-13. This result therefore implies that at most a small fraction of short GRB progenitors are BH-NS mergers. Similarly, when calibrating the true released energy, an average value of $10^{49} - 10^{50}$~erg is found, which constrain possible jet launching mechanisms (see below).

For a long time, GRB jets were treated by most of the theoretical models as being 'top-hat', namely, $\Gamma (\theta) = \Gamma_0$ for $\theta < \theta_j$, and $\Gamma (\theta) = 0$ for larger angles. This is despite the fact that numerical simulations of jets propagating through a collapsing star clearly show a more complicated internal structure \cite{MW99, ZWM03}. A possible reason for the consideration of top-hat jet is the ease of the dynamical calculations, which, in this case can mostly be done analytically. Furthermore, when considering structured jets, there is a high degree of uncertainty in the exact jet structure.

\begin{figure}
\includegraphics[width=11cm]{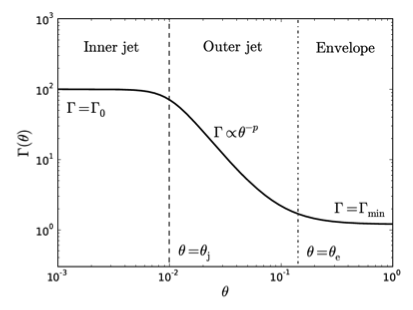}
\caption{Illustration demonstrating a lateral jet structure. The Lorentz factor is maximal in the inner jet region ($\theta < \theta_j$) and drops as a power law in angle in outer jet regions, $\theta > \theta_j$. This jet profile emerges due to shear that develops as the jet drills its way through a collapsing star \cite{ZWM03}. Figure is   taken from
  \cite{LPR13}.  }
\label{fig:jet_structure}
\end{figure}

Prior to 2017, only a handful of works considered the possible effect of a structured jet on the observed signal \cite{ZM02b, GrK03, LPR13, LPR14}. This situation had dramatically changed following the observations of GW/GRB170817. As this GRB attracted a lot of attention, high quality data exists at late times (up to months, even years). Fitting late time radio and X-ray data clearly reveled a structured jet, of the form $\Gamma(\theta) \propto \Gamma_0/\sqrt{1+ (\theta/\theta_j)^{2p}}$, namely an inner ("core") region, $\theta < \theta_j$, in which the Lorentz factor is roughly constant, and an outer, "shear" region ($\theta > \theta_j$) in which the Lorentz factor decays roughly as $\Gamma(\theta) \propto \theta^{-p}$ \cite{DAvanzo+18, Mooley+18, Troja+19}; see Figure \ref{fig:jet_structure}. Analyzing broad-band afterglow data, from radio, through optical (HST) to X-rays on time scale of months, led to the conclusion that this jet must have been structured \cite{TI20, TI21}, and viewed off axis, at an offset of $22^\circ$ from the jet axis \cite{Troja+17, Margutti+17, Alexander+18, Margutti+18, Fong+19}.  

Indeed, additional late time afterglow measurements of other GRBs, for example GRB221009A also suggest a similar structured jet \cite{OConnor+23, GG23, Zheng_J+24}. This realization of a jet structure is thus now becoming a standard when analyzing GRB signals \cite{Bosnjak+24}. Fitting data is now used to estimate the exact jet shape. 

Jet structure does not affect only the late time signal, but also the very early times, namely the prompt phase. Despite the beaming, structured jets implies that the prompt is expected to be detected even for off axis observers \cite{KBG18}. 

Of particular interest is jet structure effect on the observed signal from the photosphere, which is the earliest signal that can be detected. A jet structure 
has a major effect on the photospheric signal \cite{LPR13}, by modifying the observed spectra, both at low and high energies \cite{VP23, Vyas+24, PL24}. Various non-trivial effects, such as photon energy gain by scattering back and forth in the shear layer were discovered \cite{LPR13, Ito+13}.
The resulting spectra is far from having a "black body" shape, and rather resembles the observed "Band" function. Furthermore, it also produces a unique polarization signal (for an observer located off-axis \cite{LPR14, Parsotan+20}). A very important result is that high polarization degree is achieved from the photosphere of a structured jet, without assuming any synchrotron radiation.

To summarize, jet structure is realized in recent years to play an important role in both GRB prompt and afterglow emission phases. Studying the jet structure therefore provides a new set of constraints in studying the jet launching mechanism as well as its composition. While I focus here on the lateral jet structure, as here was a major progress in recent years, there is also a radial jet evolution that is non-trivial. For example, radio data of GRB221009A at time scale of a few days shows inconsistency with the expectation of the forward shock \cite{Laskar+23}. This suggests an additional emission component, whose origin is uncertain, as this time scale is much later than the expected for a reverse shock.

\section{Jet launching mechanism}

The initial source of energy that fuels the GRB engine is the gravitational energy of a massive star collapse, or alternatively, the merger of binary stars. The fundamental question is how this gravitational energy is converted into the form of a kinetic energy, namely to a relativistically expanding jet. Clearly, the details of the answer to this question also provide insight into the jet structure.

In the traditional "collapsar" model, the collapse of the stellar core leads to the formation of an accretion disk, rotaing around the a newly formed BH \cite{Popham+99, MacFadyen+01}. Alternatively, a millisecond pulsar (magnetar) may be formed with enough rotational energy to prevent gravitational collapse \cite{Usov92, DT92}.  In this 'proto-magnetar' model \cite{Bucciantini+08, Metzger+11, Bucciantini+12, BGM17}, the rotational energy is released as gravitational waves and electromagnetic radiation, causing the magnetar to spin down. If the magnetar is sufficiently massive it may reach a critical point at which differential rotation is no longer able to support it, resulting in collapse to a BH.

Within the original "collapsar" model, i.e., neglecting BH rotation, energy conversion is mediated by a strong flux of neutrinos, that are produced in the inner regions close to the newly formed BH \cite{Popham+99, Fryer+99, ZB11}. The neutrino -  anti-neutrino annihilate into $e^\pm$ pairs, thereby triggering the formation of the "fireball". 

An alternative scenario for jet launching is the Blandford-Znajek process \cite{BZ77}. In this model, the source of energy is the rotational energy of the newly formed BH. This energy is extracted by magnetic field lines that are brought to the horizon as they are attached to the accreting disk. They then act as 'springs', expanding by their self pressure, and convert the rotational energy into Poynting flux-dominated outflow. Particles are introduced into the jet at a later stage, e.g., by instabilities that develop at the jet boundaries  \cite{Wong+21}. These particles are accelerated, thereby converting the Poynting energy to kinetic energy, although the details of this last conversion are still uncertain. While this mechanism is in wide use in the study of AGNs and XRBs, it was only recently claimed to be highly efficient in the context of GRBs as well \cite{TG15}. 

\begin{figure}
\includegraphics[width=11cm]{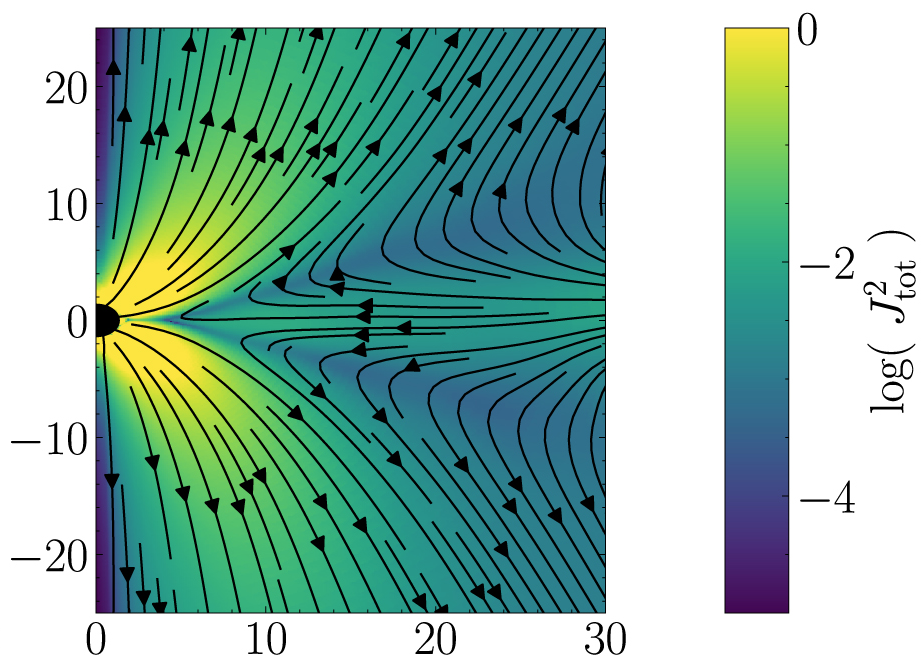}
\caption{Demonstration of a result obtained by 3D GR-MHD simulation. This one shows the flux of the $\phi$ component of the angular momentum between the disk and the jet (whose boundaries are marked by the red line). Here, the BH assumes a positive spin, $a=0.94$, and the angular momentum is the sum of angular momentum in the magnetic field and the gas. The scale is in normalized to the gravitational radius of the BH.  These results show the formation of the jet, its structure, and that the it transforms a significant amount of angular momentum to infinity.  The result is taken from \cite{Zhang_GQ+24}.  }
\label{fig:jet_production}
\end{figure}

In the past two decades, there had been a rapid development in parallel computation facilities. This enabled the developments of various codes that explore the core collapse during the stellar explosion (e.g., \cite{OA20, OA21, OA22}) as well as  general relativistic, magnetohydrodynamic (GR-MHD) codes aimed at exploring the evolution of the disks and emerging jets; as an example, see Figure \ref{fig:jet_production}. Over the years, several GR-MHD codes have been developed  (e.g., \cite{De-Villiers+03, Gammie+03, Anninos+05, DelZanna+07, Stone+08, Etienne+15, Porth+17, Liska+22, cuHARM23} and more). These codes are most frequently used in studying the properties of accretion disks around rotating BHs. Given initial conditions, the codes trace the evolution of the gas as it accretes into the BH. The numerical calculations enable to trace the various instabilities that develop, such as magneto-rotational instability (MRI; \cite{BH91, BH98}), which strongly affects the magnitude and global magnetic field configuration evolution. A major finding was that the accretion disks evolve into two distinct quasi-steady state structures. The fate of the disk evolution largely depend on the initial magnetic field configuration. 

The two quasi steady stae disk configurations are the "Standard and normal evolution" (SANE; \cite{Narayan+12}) and "Magnetically arrested disks" (MAD; \cite{IN03, Narayan+03}). These separate configurations are important, as it was found that in addition to the different disk structures, the emerging jets are much more powerful when the disks are in the MAD states \cite{Igu08, YN14}. Furthermore, 
these codes enabled detailed numerical study of the Blandford-Znajek process \cite{TNM11}, confirming its validity. 

In recent years, several authors applied some of these codes to study the properties of the emerging jets from GRBs \cite{Ito+19, Gottlieb+21, Gottlieb+22a,  Gottlieb+23, RTG24}. These GR-MHD codes do not simulate the entire collapse or merger, but rather it is assumed that the merger or collapse leads to the formation of an accretion disk surrounding a newly formed rotating BH. The simulations are then run to explore the emerging jet properties under various assumptions on the initial disk structure, magnetic field configuration, etc. These properties include, among others, jet velocity profile, fluctuations, location of internal shocks and magnetic field configuration. When added radiation, which is currently done post-processing (i.e., separated from the dynamical calculations), one also obtains the expected photospheric signal \cite{Ito+19}. It should be pointed out that deep in the flow, in regions of very high optical depth when radiation is fully coupled to the plasma, the effect of radiation can be directly incorporated by simply considering the relativistic equation of state (adiabatic index $\hat \gamma = 4/3$). 
Some authors used this to calculate the photospheric signal resulting from fluctuations deep in the flow \cite{RTG24, RTG24b}. Such calculations,
 though, are very limited, as the photon start to decouple from the gas close to, but below the photosphere \cite{Beloborodov11}, and therefore the approximations used fail.

Crudely, currently, existing simulations provide: 
\begin{enumerate}
\item Realistic structure of both the collapsing star and the newly formed disk, of course for a given set of initial conditions. There is still a high degree of uncertainty whether the initial conditions chosen represent those that occur in nature. 
\item An insight into the role of magnetic fields in the jet launching process, as well as connection between the jet properties and the inner disk properties, including the magnetic field configuration. 
\item A realistic calculation of the internal jet structure, its temporal evolution and the role of various instabilities (in particular, the Rayleigh-Taylor instability), again, for a given set of initial conditions.
\item An insight into some of the jet properties, such as its terminal Lorentz factor.
\end{enumerate} 

These results cannot be achieved analytically, and necessitate the use of very heavy numerical calculations. Therefore, existing GR-MHD codes enable a substantial progress in understanding the GRB physics. 
 
While this direction is obviously very promising, there are still very serious gaps that need to be filled before these simulations can provide realistic predictions to understand the nature of GRBs. The key gaps that still exist today include:
\begin{itemize}
\item Missing physics. Despite the great progress made, still there are important physical ingredients that are not considered in currently existing models. These include: (a) the full effects of radiation, namely radiation back reaction (i.e., its effect on the dynamics) as well as independent treatment of radiation close to the photosphere; (b) the effects of neutrinos that transfer energy, momentum and angular momentum. These transfer can be substantial under the appropriate conditions \cite{ZB11}. (c) Exact cross sections for various nuclear processes. 
\item The results of the models are sensitive to the uncertain initial conditions. This is an inherent problem that could not easily be resolved.  
\item Key ingredients, such as the initial configuration of the magnetic field, are unknown, and are therefore 'put by hand'.
\item The dynamical range of calculations is limited by computational power. Therefore, some calculations are interpolated to larger radii. As explained, when approaching the photosphere, this interpolation becomes less valid. 
\item There are various numerical limitations, such as numerical treatments of the polar regions, the need for 'flooring' (adding material 'by hand' into empty regions, in order to achieve computational stability), etc. 
\end{itemize}

Nonetheless, many of these are technical problems, that are expected to be solved in the coming years, with the continuous developments of algorithms as well as the continuous increase in computational power. I therefore anticipate that the role of GR-MHD simulations in the study of GRBs will increase in the coming years, and they would enable to provide new insights into some of the yet unsolved problems.

\section{The nature of energy dissipation mechanism}

Perhaps the easiest way to understand the complex chain of events that lead to a GRB is to follow the energy conversion episodes.
The GRB "fireball" model (which is referred to here in a very broad context) provides the basic skeleton. The source of energy is the gravitational energy, of either a collapsing star or a merger of binaries. During the formation of the BH, (part of) this energy is then converted to kinetic energy, in the form of a jet. This conversion can be mediated by neutrinos and photons (the so called "fireball"), or alternatively by magnetic field, in which case there is another energy conversion of Poynting energy to gas kinetic energy.

Since an observer does not see directly a kinetic energy, the next stage must be a mechanism that converts part of this kinetic or magnetic energy into photons. A plausible intermediate step is use of this energy to accelerate particles to high energies. These energetic particles then radiate the photons observed. Within this framework, a photospheric model provides an alternative to this part of energy conversion, as it does not require energetic particles, but rather assumes that one directly observes photons that decouple the plasma at the photosphere. 

In the early days (mid 1990's), internal energy dissipation in the form of 'internal shocks', resulting from velocity differences within the jet was suggested as a way of converting the kinetic energy to energetic particles \cite{RM94, PX94, SP97b}. This seem a natural outcome, as the flow is relativistic, and shock waves are common. Furthermore, a propagating shock wave is required to explain the afterglow.
However, it was quickly realized that the internal shocks idea suffers a severe efficiency problem, with a typical efficiency of no more than a few percents \cite{Kobayashi+97, DM98, Beloborodov00, Spada+00, GSW01}. This is due to the fact that only the differential kinetic energy is available for extraction. 

This severe drawback motivated the search for alternatives. A leading alternative that also had been discussed since the 1990's is that of magnetized jet. 
In a magnetically-dominated flow, magnetic field lines of different orientations reconnect, thereby releasing a magnetic energy that is used in heating and accelerating particles. While the basic theory of magnetic reconnection was studied already in the 1950's and 1960's \cite{Parker1957, Sweet1958, Petschek64}, the original theoretical models showed that this process may be too slow to be relevant to GRBs, making this idea less appealing until the last decade. 

The rapid development in parallel computational facilities, enabled a rapid progress in this field as well. Studying magnetic reconnection is done using 
particle-in-cell (PIC) simulations \cite{SS12, SS14, WU17}. These simulations trace the evolution of individual charged particles along a grid, by solving for the electromagnetic forces they exert on each other. Inside each grid cell, the currents resulting from particle motion are calculated. Using these currents, electromagnetic (EM) fields on the grid are
computed by summing over all the particles in a cell. These EM fields are then interpolated back to the particles in each cell, from which the Lorentz force acting on individual particles is deduced. The particles motion are then calculated from the Lorentz force.

\begin{figure}
\includegraphics[width=10cm]{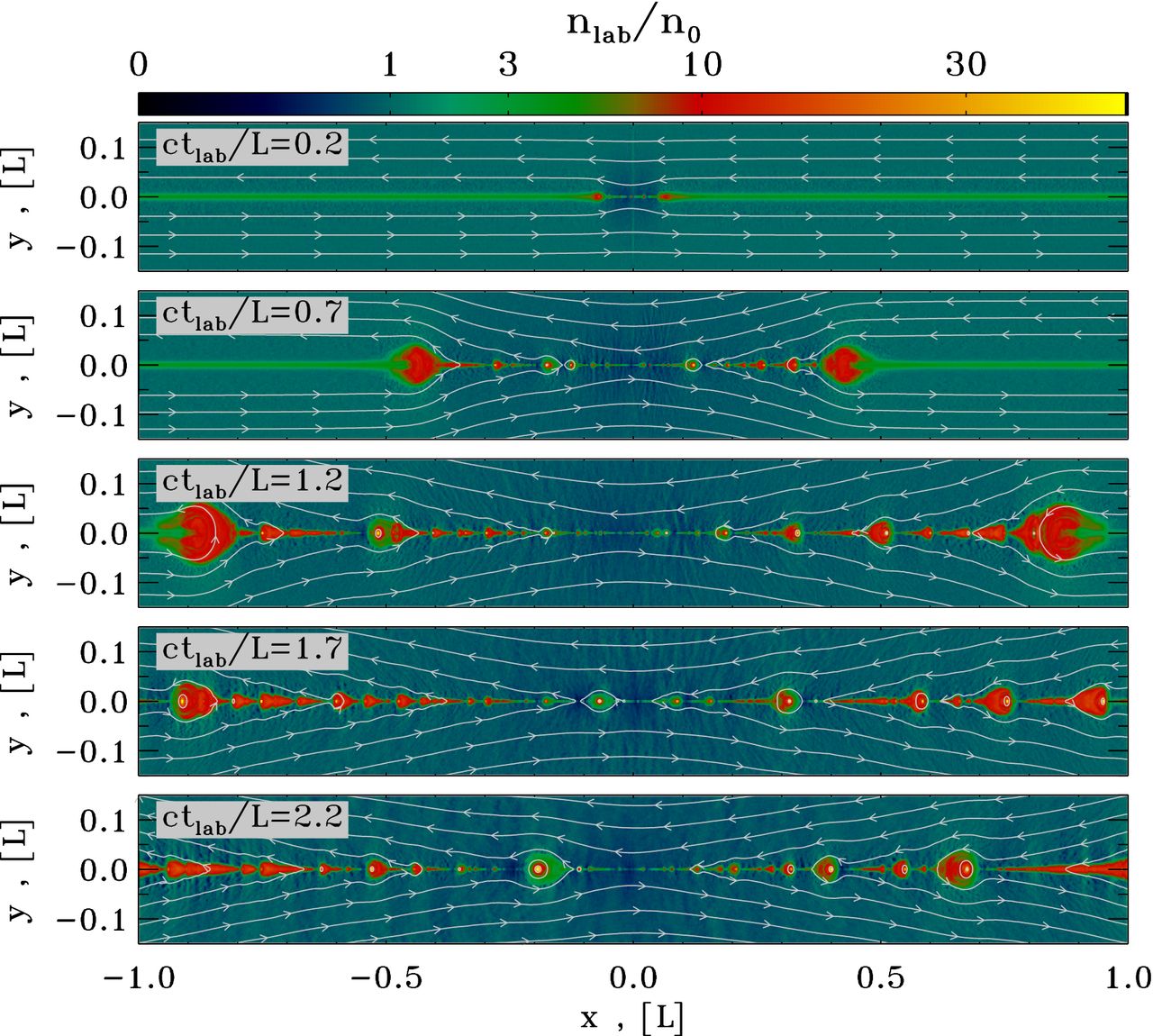}
\caption{Results of 2D simulation shows the partcle number density along the magnetic reconnection layer. Time evolves from top to bottom, as marked.  After triggering reconnection in the centre of the current sheet (x=0 in the top panel), two ‘reconnection fronts’ propagate to the right and to the left. The result is taken from \cite{Sironi+16}.  }
\label{fig:reconnection}
\end{figure}

These simulations
matured in the past decade, and provide an insight into the mechanism of magnetic reconnection.
It was found that the reconnecting lines lead to the formation of plasmoids, which are regions in space filled with energetic particles and magnetic fields \cite{Sironi+16}; see Figure \ref{fig:reconnection}. Particles are accelerated by generated electromagnetic potential, and can reach substantial energies as they enter  'reconnection island', and are then accelerated by strong electric fields that are formed between the islands \cite{Sironi+16, Sironi+23}. The limit occurs when the particle's Larmor radius becomes comparable to the plasmoid size. The accelerated particles leave the plasmoid due to the developed turbulence, and their emerging distribution is a power law \cite{ZSG21, ZSGP23}. 
The plasmoids themselves grows with time, and can reach a substantial fraction of the system size. Furthermore, due to turbulence in the flow, the rate of reconnection can be much higher than initially thought \cite{CS18, CS22, CJ23}.

Thus, overall, the rapid progress in this field in the past 10 years puts reconnection as a very viable method for explaining particle acceleration. I anticipate here too a rapid progress in the coming years.

\section{The ambient medium: deviation from self-similar motion}

It is surprising how little is known about the surrounding medium into which GRBs explode. This is mainly due to two reasons: (1) GRBs reside in distant galaxies, which cannot be resolved directly; and (2) theoretically, little is known about the final  stages of massive stellar evolution, prior to their terminal collapse. This is the stage in which they may emit strong stellar winds, which will affect the GRB environment. It is therefore difficult to theoretically predict the stellar environment, which is affected by the stellar wind. 

Early models of GRB afterglows \cite{WRM97, WG99} show a broad band spectral distribution and late time (hours onward) temporal evolution that are well fitted with the basic theoretical model of self-similar motion \cite{BM76}. The basic idea is that following an initial acceleration and coasting phases, the relativistic GRB blast wave propagates into the ambient media in a self-similar way, namely its Lorentz factor is a power law in radius, $\Gamma(r) \propto r^{-\alpha}$, where the index $\alpha$ depends on the ambient density profile ($\alpha = 3/2$ for constant density ISM, and $\alpha = 1/2$ for a decaying density, $n(r) \propto r^{-2}$, as is expected for a constant velocity stellar wind). 

Electrons are accelerated into a power law in this propagating shock wave. 
The spectra are fitted with synchrotron emission from a power law accelerated particles \cite{RL79, SPN98}, while the lightcure evolves according to the expectation from a relativistic blast wave explosion into a constant density enironment \cite{BM76}. While initially only explosions into constant density ISM were considered, extensions were quickly made to a power law density environment, namely $n(r) \propto r^{-2}$, as is expected if the star emits a constant-velocity wind for a substantial duration prior to its terminal explosion \cite{GS02}. 

Thus, this model predicts an early time light curve fluctuations, expected either before or during the transition to the self-similar phase, while late time smooth afterglow. During the transition, a reverse shock is expected, that propagates into the plasma and wipes out the memory of the initial explosion \cite{RM92, SP99}. The time scale of the existence of the reverse shock is expected to be of the orders of 10s of seconds to minutes, i.e., close to the end of the prompt phase. It was used to explain some rebrightening seen, e.g., in GRB180720B \cite{Arimoto+24}.

In recent years, analysis of late time afterglow data reveals that the expectation for a self similar motion is not always fulfilled. Various wiggles and fluctuations are detected, that are not expected. For example, it was argued that a reverse shock may exist in the lightcurve of  GRB181201A, 3.9 days after the explosion, i.e., 3-4 orders of magnitude later than expected \cite{Laskar+19b}. A second example is GRB201216C, in which radio data after $\sim$ a month requires a different emission component than the forward shock \cite{Rhodes+22, Abe+24}.
Another peculiar event was the  short GRB210726A. No radio signal was detected during the first 11 days, but then the radio flux showed a re-brightening by a factor of 3 over a duration of a week \cite{Schroeder+24}. Such a result cannot be explained as due to the forward shock, and explaining it requires either a very late reverse shock, or late time energy injection. These results, therefore, challenge the basic self-similar motion picture.

The environment profile in the stellar vicinity is expected to be much more complicated than the power law description often assumed. As the stellar wind from a GRB progenitor star is emitted over a finite, uncertain duration of thousands to millions of years, it cannot cover the entire relevant space. Instead, the massive star that emits the wind is surrounded by a "wind bubble" structure \cite{Weaver76}.

\begin{figure}
\includegraphics[width=11cm]{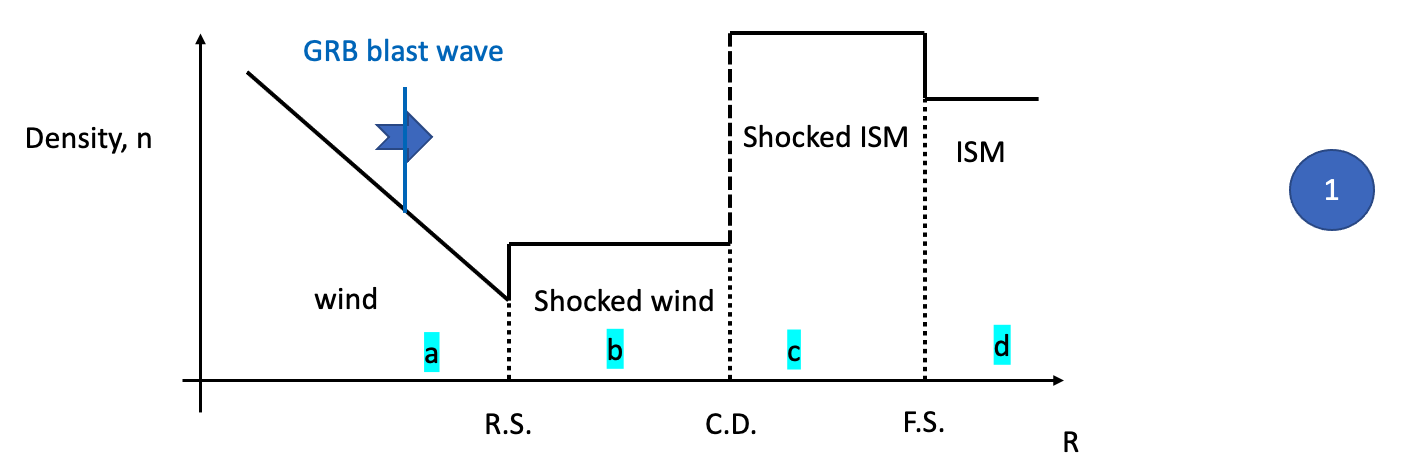}
\caption{An illustration demonstrating the four regimes in the wind bubble. The star is to the left, emitting a wind prior to its explosion. Region $a$ is the unshocked wind. Region $d$ is the (constant density) ISM. Region $b$ is the shocked wind, and region $c$ is the shocked ISM. When the star explodes into a GRB, the GRB blast wave propagates into this environment, interacting with the shock waves and contact discontinuity in it. Figure is taken from
  \cite{PR24}.  }
\label{fig:wind}
\end{figure}

This structure is characterized by four distinct regimes (see Figure \ref{fig:wind}). 
The inner most regime ("region $a$") contains the freely expanding stellar wind. The outer most regime ("region $d$") contains the interstellar medium (ISM). 
Two more regions are the shocked stellar wind ("region $b$"), consists of stellar wind shocked by the wind termination shock (reverse shock); and shocked ISM ("region $c$") shocked by the propagating forward shock, which also marks the edge of the bubble. The shocked wind and shocked ISM (regions ($b$) and ($c$)) are separated by a contact discontinuity; see Figure \ref{fig:wind}. 

The size of this cavity is $\sim 1$~pc, and it clearly depends on the uncertain model parameters, such as the wind velocity, the mass ejection rate and the time the star emits the wind \cite{Chrimes+22}.
 While in the basic picture the relevant radii can be calculated analytically, clearly, additional physical ingredients such as stellar rotation will further complicate the structure \cite{VanMarle+06, VanMarle+08}. 
Indeed, such ring nebulae are observed around 1/3 of the massive stars in our galaxy, in their Wolf-Rayet phase \cite{Marston97, Crowther07, Lau+22}. 

When the star explodes to produce a GRB, the GRB jet must cross the surrounding bubble \cite{PW06}. During its crossing, it encounters the reverse and forward shock waves as well as the contact discontinuity. These encounteres lead to obserable signals \cite{PR24}. For plausible wind bubble parameters, interaction of the GRB blast wave with the wind termination (reverse) shock is observed on a time scale of a few seconds, and may therefore be associated with an observed precursor. The main interaction may take place with the contact discontinuity, at observed time of the order of $\sim 100$~s. This could lead to a significant observed signal, which will be detected as a strong re-brightening at this time frame. Energy dissipation at this stage is much more efficient than internal shocks, as the contact discontinuity is nearly static. 

This model can explain       
about 5-10\% of GRB lightcurve that show a weak precursor, followed by a quiescent period and a main emission after 100-200~s \cite{Zhu15, Coppin+20}. 
As a concrete example, the bright GRB160821A \cite{Vidushi+19, Ravasio+19, Ryde+22}
 showed a giant flare at $\sim 100$~s, which is not directly connected to the following "afterglow" emission. 
This could very well be due to the blast wave - bubble interaction. 
One can conclude therefore that observations at the time scale of $\sim$minutes may provide new insights on the wind structure in the vicinity of GRB progenitors, hence may be used as an independent probe of the last stages in the life of massive stars that end their lives as GRBs.

\section{The X-ray plateau: potential revolution}

Associated with the question of the environment, is the open question of the origin of X-ray plateau. The plateau in GRB X-ray lightcurves was identified shortly after the launch of the {\it Swift} mission \cite{Zhang+06, Nousek+06}. Prior to {\it Swift}, data existed only during the prompt phase, and then during the later afterglow (after time scale of $\sim$~hour). {\it Swift} bridged this gap, by enabling a near continuous probe of the X-ray afterglow from the prompt phase onward. The surprising result is that immediately after the prompt phase, the flow does not transform into a similar motion as is expected \cite{BM76}. Rather, the X-ray lightcurve in a significant fraction, of about 60\% of GRBs \cite{Srini+20} is flat, for a long duration of several hundreds to several thousands of seconds - tens of minutes, sometimes even a few hours.

Over the years, a plethora of ideas were suggested to explain this result. Notable suggestions include the following. (1) A continuous energy injection that slows the acceleration \cite{Zhang+06, Nousek+06, FP06, Granot+06, Ghisellini+07}. This requires the GRB progenitor to operate for a much longer period than a few seconds. (2) Emission in inhomogeneous media \cite{Toma+06, Ioka+06}, that causes rebrightening of the lightcurve. (3) Dominant emission from a long-lasting reverse shock \cite{UB07, Genet+07, Hascoet+14}. This idea requires that the (microphysical) properties of the plasma shocked by the reverse shock will be significantly different than at the plasma shocked by the forward shock. (4) Jets viewed off axis, namely a viewing angle effect \cite{EG06, Oganesyan+20}. This is particularly appealing for structured jet viewed off axis, when gradually inner, brighter regions become accessible \cite{Ascenzi+20, Beniamini+20}. Finally, (5) emission during the coasting phase \cite{ShM12, Dereli+22}. As was found, if emission occurs into a low density "wind" environment ($n(r) \propto r^{-2}$) during the coasting phase, the resulting lightcurve can be flat \cite{Dereli+22}. The fact that no consensus had been reached after nearly 20 years since its discovery, implies that this is still considered an open question, which is debated in the literature.

\begin{figure}
\includegraphics[width=6.5cm]{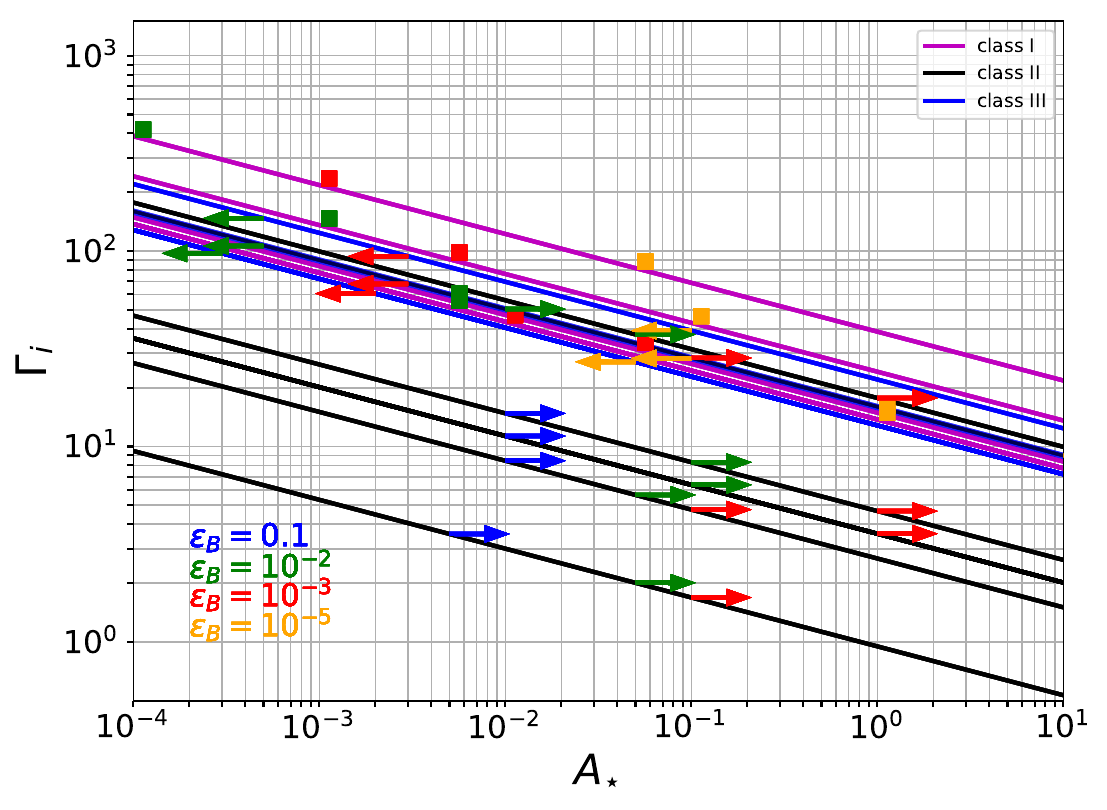}
\includegraphics[width=7cm]{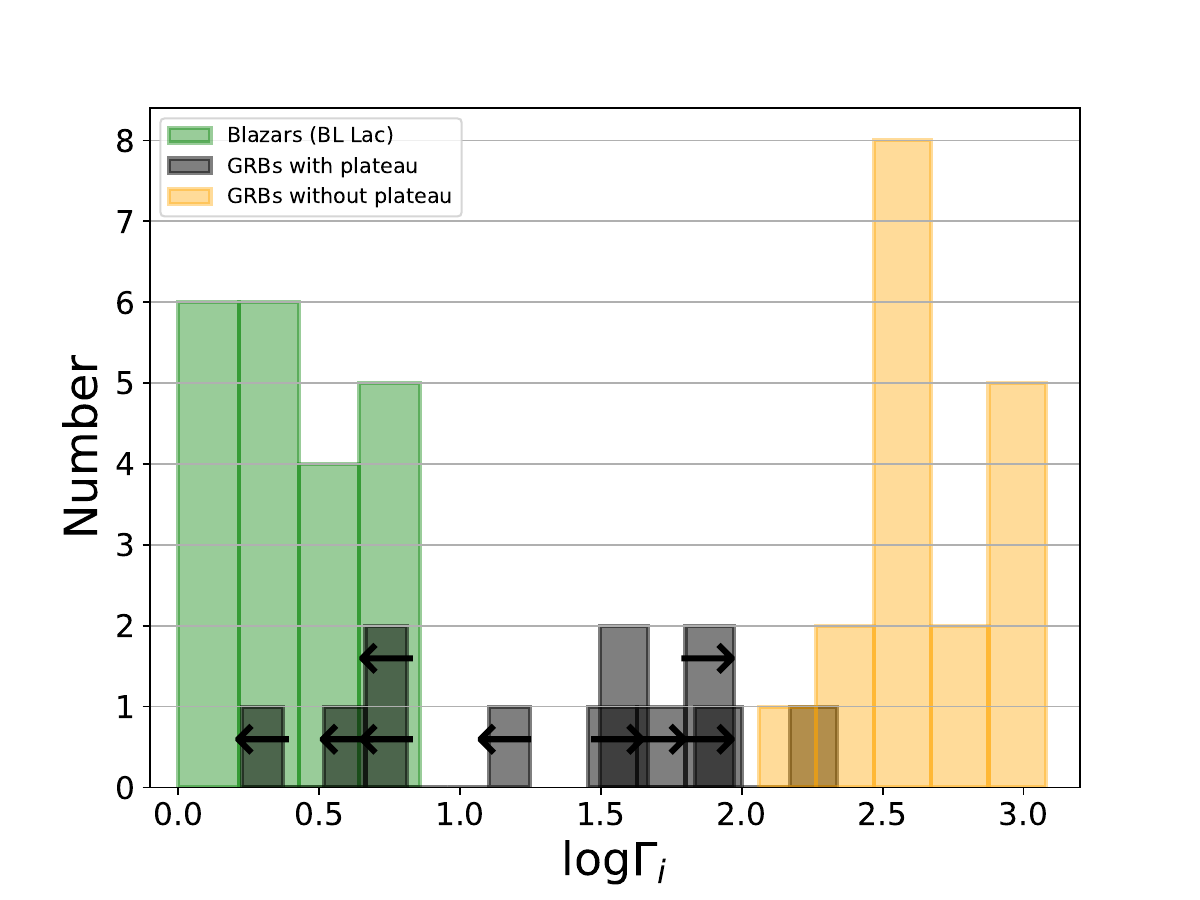}
\caption{The results of an analysis shows that the plateau can naturally be explained as due to emission during the coasting phase of jets with mild Lorentz factors, propagating into low density stellar wind. The left panel shows the constraints set on the Lorentz factor and the wind density (marked as $A_\star$, where $A_\star =1$ is expected for a Wolf-Rayet star). The right panel compares the results obtained (black) to measured Lorentz factors in AGNs (a few, marked in green) and high GRBs without plateau (several hundreds, yellow). The region of Lorentz factor in GRBs that show a plateau, of several tens, therefore fills the observational gap. The figure is taken from
  \cite{Dereli+22}.  }
\label{fig:plateau}
\end{figure}

The last idea - emission during the coasting phase, may hold the key to revolutionizing our understanding of GRB jet physics. The reason is that it was proven  
that a plateau emission is a natural outcome of a model in which the Lorentz factor of the flow is only a few tens, rather than a few hundreds, as is often assumed \cite{Dereli+22}. 
The average Lorentz factor of GRBs in the analyzed sample in that work is $\langle \Gamma \rangle \simeq 50$, with variations between a few and a couple of hundreds (see Figure \ref{fig:plateau}).

The reasoning behind the claim that GRB Lorentz factors reach terminal values of  several hundreds are as follows. (i) The opacity argument: photons with energies that exceed the threshold energy (0.5 MeV) will produce $e^\pm$ pairs \cite{KP91, WL95, LS01}, unless the observed signal is highly blue-shifted. (ii) Identifying the onset of the self-similar motion by identifying emission from the reverse shock \cite{MR97, KZ03}. The observed time is related to the terminal value of the Lorentz factor. (iii) Deducing the value of the Lorentz factor directly from measuring the properties of the thermal emission component \cite{Peer+07, ZP10, Hascoet+13}. 

A close analysis reveals that none of these observational constraints apply to GRBs with X-ray plateau. Only 3/186 GRBs in the Fermi LAT catalog \cite{Ajello+19} show any evidence for a plateau, implying an anti-correlation between the existence of a plateau and high energy emission. Furthermore, no evidence for a substantial thermal emission component, and no clear identification of reverse shock emission were observed in GRBs with plateau. These results therefore suggest that the distribution of terminal Lorentz factors within the GRB population may be much broader than previously assumed, ranging between a few (say, $\Gamma \sim 10$) to several hundreds. 

In the past year, several supporting evidence for this idea were found. 
These result from analyzing prompt emission pulses \cite{Gowri+24}, from analyzing GRB spectral lags \cite{Vyas+24}, and, most importantly, from analyzing the properties of the observed late time X-ray flares \cite{Dereli+24}. This last analysis is of particular importance, since different explanations for the origin of plateau give different, testable theoretical predictions. For example, if the plateau originates from observers located off the jet axis, then the observed time of the X-ray flares are expected to be later than for GRBs without plateau due to the different Doppler boost. The results of the analysis show that there is no difference between the average flare times for GRBs with and without plateaus, which seems to contradict this prediction. This, though, is expected if the Lorentz factor of GRBs with plateaus is lower, since in this case the flare emission radii is smaller, and the dependence on the Lorentz factor cancels.

This idea of low Lorentz factor GRBs, if proven correct, obviously marks a paradigm shift in the study of GRBs, by proving that the majority of GRBs in fact have Lorentz factor of tens rather than hundreds.  
One can conclude that this epoch of early afterglow provides several open questions that are still unanswered, and I anticipate that it will continue to be explored in the coming decade. It had already showed the potential of revolutionizing our understanding of GRB physics, if indeed proven that the Lorentz factor of many GRBs is "only" of few tens, as recently suggested.

\section{Radiative processes and radiative counterparts}

Since nearly the entire signal detected from GRBs is electromagnetic, the basic question of its origin is a fundamental one. The nature of the observed spectra strongly depends on the radius in which it originates. Emission from small radii- i.e., below or close to the photosphere is expected to be a modified black body, with a leading radiative process of inverse Compton (IC) scattering \cite{PMR06, Peer08, PR11}. On the other hand, emission at larger radii - above the photosphere, is expected to be mainly of synchrotron origin, and IC scattering contributing to the high energy part.

Already in the early 1990's when it was realized that the observed GRB spectra does not resemble a "Planck" function, synchrotron emission was suggested as a leading radiative mechanism \cite{RM94, Tavani96, Cohen+97}. However, inconsistency with the synchrotron model prediction \cite{Preece+98} (but see \cite{Burgess+20} for a reanalysis) prompted interest in alternative models. One example is a revived interest in proton-synchrotron model that was found to better fit the observed spectral slopes \cite{Ghisellini+20}; see further discussion below.
Alternatively,
a photospheric (thermal) model contribution become the subject of increasing interest 
(\cite{MRRZ02, Ryde05, PMR06, Giannios06, Ryde+10, Guiriec+11} and many more).

It should be noted that since the photospheric radius strongly depends on the Lorentz factor, $r_{ph} \propto \Gamma^{-3}$ (e.g., \cite{RM05, Peer+12}), a low Lorentz factor implies a larger photospheric radius, resulting in a likely more pronounced contribution from the photosphere. This, though, will necessitate sub-photospheric energy dissipation, or alternatively, lateral jet confinement, to reduce adiabatic losses that will lower the peak energy below the sub-MeV range in which it is observed.

Although initially, emission from the photosphere was expected to resemble a "Planck" function \cite{Goodman86}, it was realized that this approach is too simplified. There are several effects that act to broaden the naively expected black-body signal from the photosphere. First, sub-photospheric energy dissipation that occurs due to any cause- being shocks, magnetic reconnection, neutrino annihilation, or any unspecified dissipation process, will modify the emitted spectra. If the dissipation does not occur too far below the photosphere, the photons will not have sufficient time to re-thermalize, and the resulting spectra will be broadened \cite{PMR05, PMR06, Giannios06, VLP13, Ito+14, GLN19}. Second, due to the relativistic motion of the jet, light aberration effects will further modify the observed signal \cite{Peer08, Beloborodov10, PR11}. This will become very pronounced for any non-spherical expansion, such as a structured jet \cite{LPR13, LPR14, VP23, Vyas+24} - i.e., a jet with an angle-dependent Lorentz factor, $\Gamma = \Gamma(\theta)$, which is expected in a realistic scenario, as discussed above. Third, the photospheric signal, like any signal in a transient event, is time dependent. Therefore, integrating over a finite time automatically smears the signal.  And fourth, there are instrumental effects- due to the limited bandwidth of the detectors, as well as the "curvature" of the spectrum near the energy peak, it is found that the expected observed values of the low energy spectral slope are much shallower than the Rayleigh-Jeans slope \cite{Acuner+19}.

After considering these effects, a recent analysis shows that in fact more than 1/4 of long GRBs, and 1/3 of short GRBs are consistent with having a pure thermal origin \cite{Acuner+20, DPR20}. When adding a possible sub-photospheric energy dissipation that can potentially broaden the spectra, these fractions get much larger, close to 100\%.

On the other hand, a recent analysis of time-resolved GRB spectra of single pulse GRBs, showed that synchrotron emission can account for about 95\% of the spectra \cite{Burgess+20}. Thus, overall, the debate on the radiative origin of the observed signal is still on-going, and may potentially be resolved in the next decade, with a more refined time-resolved analysis.

\subsection{The pair annihilation line in the BOAT GRB 221009A: further constraints on the physical parameters}

Existence of energetic photons implies that a large number of $e^\pm$ pairs are expected to be produced within a GRB outflow. Indeed, as discussed above, within the framework of the GRB "fireball" model, existence of these pairs is a natural outcome. These pairs, in turn, are expected to annihilate, producing a distinct line, at observed energy $\Gamma m_e c^2$, where $\Gamma$ is the outflow Lorentz factor. Detection of such a line had long been predicted \cite{PW04}. 
It was therefore a surprise that such a line was never detected. A possible reason for that is the lower sensitivity of existing detectors in the $>$~MeV band.

This situation had changed recently, with the observation of GRB221009A- the "brightest of all times" (BOAT) GRB \cite{Burns+23}. This burst was so bright, that its observed fluence,  $0.21 \pm 0.02~{\rm erg~cm^{-2}}$ (as seen by Konus-Wind; \cite{Frederiks+23}) was more than 50 times larger than that of the second brightest GRB observed to date, GRB230307A.

In addition to being so extremely bright, this GRB showed clear
evidence of an emission line, starting approximately 80 seconds after
the onset of the afterglow (at 226~s after the explosion
\cite{Ravasio+23}). This line was detected at $\sim 10$~MeV, and its
peak energy showed a clear decay in time, as $\epsilon_{peak}(t) \propto
t^{-1}$. Such a discriminated line can result from the annihilation of
pairs. However, a direct calculation results in Lorentz factor of
$\sim 20$ (the Doppler boost needed to reach this energy), which seem too low given the extreme brightness of this GRB.

A more comprehensive calculation carried recently \cite{PZ24}, examined the conditions for producing such a line. 
The temporal decay of the peak is a strong hint towards high latitude
emission, i.e., that the emission at different times
originates from off the line of sight,
therefore the Doppler boost varies with observed time as an observer sees emission from different angles \cite{KuPa00}. Taking this into account,
it was shown that this GRB jet has a more realistic Lorentz factor, of $\Gamma \approx 600$. Most importantly, a detailed analysis revealed that only a relatively narrow range of physical conditions: very high luminosity and Lorentz factor that is in a relatively narrow range of few hundreds is needed in order to produce the observed pair annihilation line signal. 
The conditions found, in a range that is much narrower than previously thought, explains the rareness of this line (see \cite{PZ24} for further details). These results therefore demonstrate that further identifications of such annihilation lines can be very useful tools in constraining the physical properties of GRB jet outflows.

\subsection{TeV emission and its origin}

Another field which matured in the past decade is that of very high energy detectors. 
In the past decade, we witness the matureness of high energy (GeV- TeV range) detectors, such as
MAGIC \cite{LM05, Melandri+22}, H.E.S.S \cite{Abramowski+14, HESS21} and recently
LHAASO \cite{LHAASO19, LHAASO23, LHAASO23b}.
For example, the MAGIC collaboration published recently
lightcurve and spectra of GRB190114C \cite{Acciari+19, Ravasio+19b, Melandri+22},
starting about a minute after the onset of this burst, and lasting for
about 40 minutes. MAGIC data shows a comparable flux at the TeV band
to that of longer wavelength, in particular the GeV (Fermi-LAT band)
and X-rays (XRT and Fermi-GBM bands). Similarly, LHAASO reported 7 / 13 TeV photons in GRB221009A  \cite{LHAASO23b, LHAASO23c}.

These new data naturally called for a theoretical interpretation. A
basic model that was suggested as a way of explaining the TeV data is
that of IC scattering (e.g., \cite{Acciari+19, DP21}). Indeed, this process is naturally expected: as
energetic electrons are needed to explain the lower energy (optical, X- and
gamma-rays) signal observed by synchrotron emission, they must be
accelerated to high energies inside the plasma. These energetic
electrons up-scatter the synchrotron photons to
higher energies, and may therefore contribute to the TeV signal. Additional
advantages of this model is that the energy budget needed is
relatively not very high, and the required magnetic field energy is
relatively low.

However, a close look reveals that this model requires some additional
assumptions in order to provide good fits to the TeV data. Both the
flux and the observed spectral and temporal slopes predicted do not
match very well those observed. In order to overcome these problems, additional
'freedom factors' were suggested \cite{DP21}, which enable some fine tuning of the
model parameters. Such freedom parameters include a certain freedom
factor in connecting the emission radius and the observed time, the
emitted and observed frequency or the dependence of the observed
luminosity on the jet kinetic energy; it turned out that the use of
the classical, basic theoretical relations do not provide sufficient
fits.

An alternative model that was proposed is that of synchrotron emission
from accelerated protons \cite{Isravel+23a, Isravel+23b}. This idea is not new, as  similar ideas were already proposed in the 1990's
(e.g., \cite{BD98, Totani98, Razzaque+10}). These were less appealing
due to the fact that protons are much less efficient radiators than
electrons. As a result, total high energy budget is required, needed
to be provided to the energetic protons to reproduce the observed
flux.

However, as pointed out recently \cite{Isravel+23a}, 
the problem can be easily overcome by noting that 
it is sufficient to assume that only a small fraction, of $\approx
10\%$ of the protons are accelerated. This is both aligned with the
results of particle-in-cell simulations, that show that only a few \%
of the particles are accelerated in shock waves \cite{SS11}, and lead
to a dramatic decrease in the required overall explosion energy budget, which is
$\sim 10^{54.5}$~erg - high, but not unreasonable.

According to this model, both electrons and protons are accelerated by
the propagating shock. The electrons though have a lower energy, and
therefore are in the slow cooling regime, while the more energetic
protons are in the fast cooling regime \cite{SPN98}. 
Radiation in the X-ray and gamma-ray band is
explained by synchrotron emission from the electrons, while the TeV
emission is due to synchrotron emission from the accelerated
protons. The condition for proton-synchrotron to dominate over IC
scattering is that the fraction of post-shock thermal energy converted
to the magnetic field is much higher than the fraction of energy given
to the electrons, namely, $\epsilon_B \gg \epsilon_e$. Here, $\epsilon_B$ is the fraction of post-shock thermal energy that is converted to magnetic field, and $\epsilon_e$ is that fraction used in accelerating electrons above the thermal distribution. Indeed values
that are found to fit the broad band data of GRB190114C are $\epsilon_B = 0.13$ and $\epsilon_e
= 0.003$. Both values are consistent with the results of PIC
simulations, as well as with fits of late time afterglow data of
various GRBs. In the opposite regime, $\epsilon_e \gg \epsilon_B$, IC
emission from the electrons is found to dominate proton-synchrotron
contribution.  These results seem to be universal: similar fitting
holds also for the spectra and lightcurve of GRB221009A \cite{Isravel+23b}. Similarly, protons accelerated at the reverse shock may contribute as well to the TeV flux \cite{Zhang_T+23, Zhang_T+23b}.

Thus, continuous stream of TeV data, as is expected with the matureness of current TeV detectors, and the coming CTA observatory \cite{CTA19} may revolutionize our understanding of the radiative processes, and of the protons role in the observed signal. This will clearly have a direct impact on the physics of cosmic rays and expected high energy neutrinos.

\subsection{Polarization: introducing a new dimension}

The final signal that I would like to mention, is that of
X- and $\gamma$-ray polarization. These represent another observational field that is maturing in recent years. 
While early claims of a high degree of polarization from GRBs exist for over 20 years \cite{CB03, Yonetoku+11}, these were by large sparse and not always reliable. Polarization measures were expected, though, as both leading radiative models in GRBs, namely synchrotron emission and Compton scattering are predicted to produce a high degree of polarization \cite{GW99, GrK03, Lyutikov+03, LPR14}.

Following the launch of AstroSat, a
plethora of polarization information became available \cite{Chatt+22}. This is due to  the Cadmium-Zinc-Telluride Imager (CZTI) on board this satellite, which is sensitive at both soft and hard X-ray (0.3-100 keV). 
A unique
example is GRB160821A. This was a very energetic burst ($E \gtrsim
10^{53}$~erg), which show high degree of polarization - more than 30\%
in the gamma-ray lightcurve. The most interesting observation was
a polarization angle change, which was detected twice: once during the
rise phase and once during the decay phase of a bright pulse that was
seen after $\sim 120$~s. Each of these polarization angle changes is
consistent with 90 degrees \cite{Vidushi+19}.

Such a polarization angle change challenges existing models. While
current models can explain a 90 degrees change for an observer located
close to the jet edge (e.g., \cite{GKG21}), this is accompanied by a
sharp decrease in the flux. The reason for this flux decay is that in
order to obtain such a 90 degrees change, the observer needs to be
located close to the jet edge. Initially, detected photons originate from a small,
magnetized region, which produces a polarized signal. As time elapses,
the region from which photons reach the
observer grows, and part of it is outside of the jet. Thus, while the
observed parts that remain within the jet can produce a signal
polarized in 90 degrees to the initial polarized signal, most of the
viewing region is outside of the jet opening angle, and therefore the
flux is expected to sharply drop.

As of today, a convincing explanation to this observational result is
still lacking, and it requires some 'out of the box' ideas, such as
unconventional jet geometry. Indeed, most of existing theories for polarization assume simple, 'top-hat' jets. However, recently, more advanced models, that consider the possiblity of jet structure emerge \cite{Birenbaum+24}. I anticipate that many more such models are expected to emerge in the coming years, with the realization that GRB jets are structured.

\section{Summary}

Extensive study of GRBs began in the early 1990's, about 30 years ago. Despite the matureness of the field, basic open questions still remain. In this short review, I tried to highlight the key advances that took place in the past decade, while looking into the next decade and trying to predict the next challenges that are expected to be addressed in the coming years. 

Looking at the different subjects, one can summarize as follows. 
\begin{enumerate}
\item {\bf The nature of progenitor.} Ten years ago there were already firm evidence supporting the idea that long GRBs originate from the collapse of a massive star (the "collapsar"), and there were plenty of indirect evidence supporting the idea that short GRBs originate from a binary merger. Today, while there is a consensus that long GRBs indeed originate from a collapsar, there are several outliers known, whose origin is not clear. There is one firm detection of the association of a short GRB with a merger- the GW/GRB170817 event, but it is not fully clear whether this event is representative of the entire short GRB population. 

\item {\bf Jet launching mechanism and GRB jet composition.} Ten years ago, the basic theories - namely the Blandford-Znajek \cite{BZ77}, collapsar and merger already existed, but most of the details were uncertain. In the past decade, a major progress in computational facilities took place, which enabled to study these mechanisms in much details. These include many relevant physical processes that cannot be studied analytically, such as instabilities, effect of radiation on the dynamics, and magnetic field configurations. I anticipate that this field will continue to flourish in the next decade. 

\item{ \bf Jet structure, dynamics and evolution.} Ten years ago, most works considered a simple 'top hat' jet, i.e., a jet with a sharp cutoff, as well as 'standard' (self-similar) jet dynamics. In the past decade, and especially after GW/GRB170817, it was realized that GRB jets do have a spatial structure, which is now taken into consideration by several authors. Furthermore, the idea that many GRB jets have Lorentz factor of tens rather than 100's, although suggested relatively recently, have a strong potential to revolutionize the field.

\item {\bf Properties of the ambient medium.} Ten years ago, the vast majority of works assumed a very simple ambient density profile, either constant or decaying as a power law with radius from the progenitor, as is expected from a steady stellar wind. Only recently, the realization that massive stars are surrounded by wind nebulae, or wind bubbles which may have a significant effect on the early afterglow in (long) GRBs start to be explored more in depth. Here too, I anticipate a potential for further breakthroughs in the next decade. 

\item {\bf Energy dissipation mechanism.} While early models suggested shock waves, whose physics is well understood, as a leading kinetic energy dissipation mechanism, already ten years ago it was realized that this mechanism is not able to provide efficient enough dissipation. The matureness of computational facilities and PIC simulations in the past decade enables a detailed study and several breakthroughs in understanding the alternative mechanism, of magnetic reconnection. Here too, additional progress is anticipated in the next decade. 

\item {\bf Radiative process.} The basic radiative processes - synchrotron, inverse Compton and photosphere are known for decades, and were used since the 1990's and until today to fit most GRB spectra. However, in the past few years, with the increase of the data quality (and quantity) it was realized that many of these models are too simplified, and do not provide good enough fits to the data. This led to a renewed interest in alternative models, such as proton-synchrotron. 

Major efforts are devoted to obtain new signals, such as TeV and polarization, which are expected to flourish in the next decade, promising a wealth of new data. Furthermore, the need for a time dependent spectral analysis, as well as  abandoning of the "Band" function become more evident in the past few years. New detections, such as the pair annihilation line in GRB221009A challenge existing theories, and calls for a renewed modeling.  Thus, overall, the continuous streaming of new data promises to stimulate new ideas in the next decade and beyond.

\end{enumerate}

\funding{This work is supported by the EU via ERC consolidating grant \#773062 (O.M.J.) and by the Israel Space Agency via grant \#6766.}

\conflictsofinterest{The authors declare no conflicts of interest.} 



\begin{adjustwidth}{-\extralength}{0cm}

\reftitle{References}

\PublishersNote{}
\end{adjustwidth}
\end{document}